# Coupling enhanced sampling of the apo-receptor with template-based ligand conformers selection: Performance in pose prediction in the D3R Grand Challenge 4.

**Andrea Basciu[1], Panagiotis I. Koukos[2], Giuliano Malloci[1], Alexandre M. J. J. Bonvin[2*] and Attilio V. Vargiu[1,2*]**

[1] Dipartimento di Fisica, Università di Cagliari, Cittadella Universitaria, I-09042 Monserrato (CA), Italy
[2] Bijvoet Center for Biomolecular Research, Faculty of Science - Chemistry, Utrecht University, Padualaan 8, 3584 CH Utrecht, The Netherlands

*Correspondence to:
A. M. J. J. Bonvin: a.m.j.j.bonvin@uu.nl;
A. V. Vargiu: vargiu@dsf.unica.it

**ORCIDs**
G. Malloci: 0000-0002-5985-257X
A. M. J. J. Bonvin: 0000-0001-7369-1322
A. V. Vargiu: 0000-0003-4013-8867

## Abstract

We report the performance of our newly introduced Ensemble Docking with Enhanced sampling of pocket Shape (EDES) protocol coupled to a template-based algorithm to generate near-native ligand conformations in the 2019 iteration of the Grand Challenge organized by the D3R consortium. Using either AutoDock4.2 or HADDOCK2.2 docking programs (each software in two variants of the protocol) our method generated native-like poses among the top 5 submitted for evaluation for most of the 20 targets with similar performances. The protein selected for GC4 was the human beta-site amyloid precursor protein cleaving enzyme 1 (BACE-1), a transmembrane aspartic-acid protease. We identified at least one pose whose heavy-atoms RMSD was less than 2.5 Å from the native conformation for 16 (80%) and 17 (85%) of the twenty targets using AutoDock and HADDOCK, respectively. Dissecting the possible sources of errors revealed that: *i)* our EDES protocol (with minor modifications) was able to sample sub-ångstrom conformations for all 20 protein targets, reproducing the correct conformation of the binding site within ~1 Å RMSD; *ii)* as already shown by some of us in GC3, even in the presence of near-native protein structures, a proper selection of ligand conformers is crucial for the success of ensemble-docking calculations. Importantly, our approach performed best among the protocols exploiting only structural information of the apo protein to generate conformations of the receptor for ensemble-docking calculations.

**KEYWORDS**: molecular docking; metadynamics; EDES; HADDOCK; AutoDock; BACE-1

# Introduction

The Drug Design Data Resource (D3R) 2019 Grand Challenge is the fourth iteration (GC4) of the major docking competition organized by the D3R consortium [1–3]. The competition has two main goals: i) assessing the ability of docking algorithms to accurately predict the binding poses of a protein against a diverse set of small molecules, and ii) evaluating of the performance of binding affinity predictors.

The target in this iteration of the pose prediction assessment is the beta-site amyloid precursor protein cleaving enzyme 1 (BACE-1), a beta-secretase 1 protein [4]. BACE-1 plays an early role in Alzheimer's disease, as it is essential for the generation of the β-amyloid peptides composing the amyloid plaques which are the hallmark neuropathological lesions [5, 6]. Given its role in initiating the formation of β-amyloids, BACE-1 has been a critical target in developing therapies against the progression of Alzheimer's disease [7], as testified also by the huge number of BACE-1 protein structures deposited in the Protein Data Bank (PDB) [8] at the beginning of the challenge (>300 on September 4th 2018). Most of these structures contain putative BACE-1 inhibitors, further witnessing the tremendous potential of this target for treating Alzheimer's disease [9, 10].

Here we report the performance of a new approach for ensemble-docking [11, 12], which couples our recently proposed EDES (Ensemble-Docking with Enhanced-sampling of pocket Shape) protocol to sample holo-like and druggable conformations of proteins [13] by means of MD simulations [14, 15] with the template-based algorithm for ligand conformer generation successfully employed in the previous GC3 competition [16]. EDES is peculiar in that, regarding the search for holo-like conformations of proteins, it exploits only the experimental structure of the apo-enzyme (PDB ID 1SGZ [17]). The method enhances the sampling of druggable (prone to host ligands) conformations of a given receptor by means of metadynamics [18] simulations. Namely, EDES exploits an original set of collective variables (CVs) describing the geometry (that is, the shape and the volume) of the putative binding site(s). The method has been validated against targets undergoing very minor (single sidechain rearrangement) to very large (hinge-bending motions) conformational changes upon ligand binding [13]. In all cases, EDES was able to sample conformations of the binding site nearly identical to those occurring in the X-ray structures of the reference complexes, as well as to yield top ranked near-native docking poses, thus validating its potential as a new general approach to improve structure-based drug design.

The method for generating ligand structures [16] selects a number (here 10) of ligand conformations out of a pool of (500) conformers generated using the OpenEye Omega software [19] by searching for similar structures on the PDB based on the matching between extended Tanimoto coefficients. As such it still does require available 3D structures of related ligands bound to the receptor to be present in the PDB. It is however a first step toward fully blind docking from the apo-receptor in cases where no single structure of a related ligand would be present in the PDB.

We benchmark here for the first time in a blind docking experiment our hybrid approach in which proteins conformations obtained from enhanced sampling of the apo structure are combined with ligand conformers selected using the templated-based approach described previously. Near-native ligand poses were found for 16 (80%) and 17 (85%) of the twenty targets using AutoDock [20] and HADDOCK [21, 22], respectively. The most challenging ligands were those whose selected conformers displayed the largest deviation from the true geometry in the native complex. Importantly, our method performed best among those using conformations of the receptor generated without exploiting previous structural information of BACE-1 in complex with other ligands.

# Materials and Methods

The D3R GC4 is divided into a set of different stages, in which the participants are requested to predict the binding pose of a set of different ligands against the same receptor and to rank them and/or estimate their free energies of binding. We participated in stages 1a and 1b of the pose prediction challenge. Specifically, in stage 1a (cross-docking) we performed ensemble docking calculations on conformations of the protein and of the ligands generated by our methodology (vide infra), while in stage 1b (self-docking) the 10 conformers of each ligand were docked onto the conformations of BACE-1 extracted from the structures of their complexes with the 20 compounds in the dataset. For both challenges the participants were asked to generate a ranked set of maximum 5 poses for each ligand.

### *Data provided*

In stage 1a the only data provided by the organizers consisted of a list of 20 SMILES entries (corresponding to the 20 compounds for which the participants must predict the crystallographic pose) and of the protein sequence in FASTA format. In stage 1b the experimental structures of the receptors for all 20 BACE-1 complexes were provided, to allow the participants to re-dock each ligand on the corresponding holo-conformation of the receptor.

### *Ligand preparation*

The 20 ligands were similar in size, each containing about 35 heavy atoms. For the generation of their conformers we employed the methodology featured in [16]. The protocol makes use of ligand similarities in the form of Tanimoto coefficients and can be summarized in the following steps:

1. Identify existing highly homologous structures of the target protein bound to small molecules;
2. Discard undesirable structures (e.g. low resolution, split side chains near the binding site, covalently bound ligands in the case where the target ligand is known to bind non-covalently, etc.);
3. Calculate Tanimoto coefficients between all target and template ligands;
4. Generate up to 500 ligand conformers for all targets with OpenEye OMEGA [19];
5. Select 10 conformers for all targets by comparing the generated conformers to the structure of the template ligand with the highest Tanimoto similarity per ligand, with OpenEye ROCS (shape and colour mode) [23].

### *Binding site determination*

In order to identify the putative binding site of the protein, we used the same approach presented in [16]. Namely, we retrieved in the PDB all the structures featuring at least 95% sequence identity to the amino acid sequence provided by the organizers and having a co-crystallized ligand (other than crystallization buffer molecules); this resulted in 340 entries. We verified that the binding site was well characterized and perfectly conserved in all the structures with no missing residues in the putative pocket. Next, we used the Tanimoto metrics, as implemented in fmcsR [24] and chemmineR [25] packages to evaluate the similarity between the ligands present in the 340 entries and each of the 20 compounds provided, in order to identify a set of receptor templates featuring the most similar ligand to the compounds to be docked. Details on Tanimoto similarity measurement can be found in [16, 26]. The search for the structure featuring the most similar ligand to each of the 20 compounds resulted in 9 complex structures selected as templates (see Table S1). In addition to providing a metric to select ligand's conformers for docking calculations, these structures were used to identify the residues lining the binding site to enhance its conformational sampling. This list was built by merging all the residues within 3.5 Å from the ligand in each of the 9 complex structures selected. Since this approach generated a very large number of residues (more than 30), we kept only the most conserved ones (appearing at least in 2 structures) and, among those present only in one structure, the most buried ones (likely to interfere with ligand binding). The resulting selection is shown in Fig. 1, which clearly highlights that the chosen list of 20 residues (see Table S2) surrounds all the 20 congeneric ligands provided for this challenge.

### *Unbiased and enhanced-sampling Molecular Dynamics (MD) simulations*

In this work we applied essentially the same workflow presented in the original EDES publication (see [13] for details), with some modifications that are described in the following sections.

*Starting protein structure.* We used the package BLASTP 2.7.1+ [27] to search for protein structures homologous to the sequence provided by the organizers. We searched the PDB, setting the number of alignments (*-num_alignements*) to 1000 and the number of scoring evaluations (*-evalue*) to 10, while using the default values otherwise. We also requested the structures to have the word "BACE" in their name. With these search criteria, we identified around 300 structures, which included only 8 apo proteins. Among these, we identified as template the structure with PDB ID 1SGZ [17], which was resolved at good resolution (2 Å), did not feature any missing residue, and displayed a full overlap with the BACE-1 sequence provided by the organizers. Prior to setting up the system, the structure was further refined through the MolProbity webserver [28].

Unbiased MD. Standard all-atom MD simulations of the apo protein (herafter $MD_{apo}$) embedded in a 0.15 KCl water solution (~60.000 atoms in total) and under periodic boundary conditions were carried out using the *pmemd* module of the AMBER18 package [29]. The initial distance between the protein and the edge of the box was set to be at least 16 Å in each direction. Topology files were created for each system using the *LEaP* module of AmberTools18 starting from the apo structure with PDB ID 1SGZ. The AMBER-FB15 [30, 31] force field was used for the protein, the TIP3P-FB model was used for water, and the parameters for the ions were obtained from [32]. Long-range electrostatics was evaluated through the particle-mesh Ewald algorithm using a real-space cutoff of 12 Å and a grid spacing of 1 Å in each dimension. The van der Waals interactions were treated by a Lennard-Jones potential using a smooth cutoff (switching radius 10 Å, cutoff radius 12 Å). Multistep energy minimization with a combination of the steepest-descent and conjugate-gradient methods was carried out to relax internal constraints of the systems by gradually releasing positional restraints. Following this, the system was heated from 0 to 310 K in 10 ns of constant-pressure heating (*NPT*) using the Langevin thermostat (collision frequency of 1 $ps_{-1}$) and the Berendsen barostat. After equilibration, a production run of 1μs was performed. A time step of 2 fs was used for pre-production runs, while equilibrium MD simulations were carried out with a time step of 4 fs in the *NPT* ensemble (using a MC barostat) after hydrogen mass repartitioning [33]. Coordinates from production trajectory were saved every 100 ps.

Enhanced sampling MD. EDES aims to generate holo-like conformations of a protein exploiting only structural information on the apo counterpart. This is achieved by means of bias-exchange well-tempered metadynamics simulations [34, 35] on a set of generic CVs effectively biasing both the shape and the volume of the binding pocket. Ultimately, the method mimics induced fit rearrangements of the receptor due to ligand/protein interactions.

Metadynamics simulations were performed on the apo protein using the GROMACS 2016.5 package [36] and the PLUMED 2.3.5 plugin [37]. Simulations were started from the last conformation sampled along the pre-production step of the unbiased MD. AMBER parameters were ported to GROMACS using the *acpype* parser [38]. Following the original implementation, four CVs defined considering only the residues within the binding site were used: *i)* the radius of gyration of the binding site (hereafter $RoG_{BS}$) calculated using the *gyration* built-in function of PLUMED; *ii)* the numbers of (pseudo)contacts across three orthogonal "inertia planes" (CIPs), calculated through a switching function implemented in the *coordination* keyword of PLUMED. The inertia planes are defined as the planes orthogonal to the three principal inertia axes of the binding site and passing through its geometrical center. All non-hydrogenous atoms were considered to define the CIPs, while only backbone atoms were used to estimate $RoG_{BS}$, on which we also implemented a "windows" approach aimed to sample in a controlled manner different shapes of the binding site. Namely, we applied soft restraints at $RoG_{BS}$ values that are 7.5% higher and lower than the $RoG_{BS}$ of the X-ray apo structure ($RoG_{X\text{-}rayapo}$) and from that trajectory, corresponding to the first window, we randomly selected a conformation associated with a $RoG_{BS}$ value 5% lower than $RoG_{X\text{-}rayapo}$. This structure was used as starting point for another MD simulation (corresponding to window 2) with walls centered at ±7.5% $RoG_{X\text{-}rayapo}$ from this new center. We repeated the procedure until we generated 3 windows including the first one, centered at 9.91, 9.41, and 8.92 Å respectively (corresponding to a ~10% decrease of $RoG_{X\text{-}rayapo}$). Each replica was simulated for 100 ns, leading to 400 ns of metadynamics simulations per window; coordinates were saved every 10 ps. Note that in [13] we demonstrated that EDES is not sensitive to the exact choice of the windows parameters.

The height *w* of the Gaussian hills was set to 0.6 kcal/mol, while the widths $s_i$ of the Gaussian hills were set to 0.06, 2.6, 1.7 and 3.0 respectively for $RoG_{BS}$ and $CIP_{1,2,3}$, respectively. The bias factor for well-tempered metadynamics was set to 10. Hills were added every 2 ps, while the bias-exchange frequency was set to 20 ps. The force constants for the restraints on the $RoG_{BS}$ were set to 50 and 10 kcal·mol$^{-1}$·Å$^{-2}$ for the upper and lower walls respectively. Hereafter, we will refer to these simulations as $EDES_{3w}$.

## *Docking*

Ensemble docking calculations were performed using either AutoDock4.2 [20] or the HADDOCK2.2 webserver [21, 22], following the same procedures described in [13]. However, at odd with the original implementation, calculations were here performed on 200 protein conformations obtained by merging structures extracted from both $MD_{apo}$ and $EDES_{3w}$ simulations. This approach appeared to be reasonable, as the extent of the conformational changes occurring at the binding site was unknown *a priori*. In particular, since we expected to observe significant oscillations of this region including the flap (Fig. 1) also along $MD_{apo}$, including structures extracted from that trajectory should encompass a non-negligible fraction of structures with $RoG_{BS}$ larger than the value calculated on 1SGZ. In addition, the number of protein conformations was lowered to 200 to cope with the time constraints of the challenge, also considering that for each compound, 10 ligand conformations were employed in ensemble-docking runs. Protein conformations were extracted by means of a multi-step cluster analysis as described in [13], with the additional requirement to extract at least 10 cluster representatives from each of the 10 slices in which the $RoG_{BS}$ distributions were binned, so as to include a certain number of structures also from poorly sampled regions. The multi-step cluster analysis was applied separately to $MD_{apo}$ and $EDES_{3w}$, extracting 500 clusters from each trajectory. Next, an additional cluster analysis using the same approach was performed on the pool of 1000 cluster representatives in order to generate the final ensemble of 200 conformations. For AutoDock, the docking was repeated with each selected conformer, while for HADDOCK the ensemble of conformers was submitted and used in a single docking run. We submitted four sets of protein-ligand binding pose predictions generated through the following docking protocols:

1. *Autodock (receipt ID pe6zg)*: Docking calculations and pose selection were performed following prescriptions detailed for Autodock in [13]. Briefly, for each ligand, 10 different conformers were docked on each of the 200 protein structures using the Lamarckian Genetic Algorithm (LGA). The grid density and number of energy evaluations were both increased from default values (respectively by decreasing the *spacing* parameter from 0.375 Å to 0.25 and by increasing the *ga_num_evals* parameter by a factor of 10) in order to avoid repeating each calculation several times to obtain converged results. An adaptive grid was used, enclosing all of the residues belonging to the binding site in each different protein conformation. Next, the top poses (in total 200, one for each docking run) were clustered using the *cpptraj* module of AmberTools18 with a hierarchical agglomerative algorithm. Namely, after structural alignment of the protein binding site conformations, the poses were clustered using a distance RMSD (dRMSD) cutoff $d_c=0.075 \cdot N_{nh}$, where $N_{nh}$ is the number of non-hydrogenous atoms of the ligand. This choice was made in order to tune the cutoff to the molecular size of each compound. Finally, clusters were ordered according to the top score (lowest binding free energy) within each cluster.
2. *HADDOCK (receipt ID kmtri)*: Docking calculations and pose selection were performed following prescriptions detailed for HADDOCK in [13]. A single docking run was performed per case, starting from the various ensembles of 200 conformations, with increased sampling (10000/400/400 models for it0, it1 and wat steps, respectively referring to rigid-body docking, semiflexible and final refinement in explicit solvent). The weight of the intermolecular van der Waals energy used in it0 was increased to 1.0 (from the default value of 0.01), and RMSD-based clustering was selected with a cutoff of 1 Å. Docking was guided by ambiguous distance restraints defined for the residues of the binding site and the ligand. During it0 the protein binding site residues were defined as "active", effectively drawing the ligand into the binding site without restraining its orientation. For the subsequent stages only the ligand was active, improving its exploration of the binding site while maintaining at least one contact with its residues.

3. *Autodock with pose refinement and rescoring (herafter Autodock$_{rr}$; receipt ID nstab)*: This approach is the same as in 1, with an additional step consisting in the relaxation of the top 10 docking poses by means of a multi-step structural relaxation performed with AMBER18 [29]. Systems were optimized *in vacuum* through three consecutive cycles of restrained structural relaxation (1000 cycles of steepest descent followed by up to 24000 cycles of conjugate gradients) followed by an unrestrained optimization (2000 cycles of steepest descent followed by up to 8000 cycles of conjugate gradients). During restrained relaxation harmonic forces of 0.3, 0.2, and 0.1 kcal·mol$^{-1}$·Å$^{-1}$ respectively for the first, second and third cycles were applied on all non-hydrogenous atoms of the system. Long-range electrostatics was evaluated directly using a cutoff of 99 Å, as for the Lennard-Jones potential. The AMBER-FB15 [30, 31] force field was used for the protein, while the parameters of the ligands were derived from the GAFF force field [39] using the *antechamber* module of AmberTools18. In particular, bond-charge corrections (bcc) charges were assigned to ligand atoms following structural relaxation under the Austin Model 1 (AM1) approximation. Finally, the poses were rescored using the same scoring function of AutoDock employed to rank the original docking poses.
4. *HADDOCK$_{all\text{-}Hs}$ (receipt ID apue7)*: this approach is the same as in 2, except for the inclusion of all hydrogens (and not only the polar ones) in atomic models.

In addition to ensemble-docking calculations using receptor structures generated *in silico*, during stage 1b we also performed self-docking calculations (only with AutoDock), that is using for each ligand the conformation of the receptor extracted from the corresponding holo experimental structure released by the organizers at the beginning of stage 1b (protocol *Autodock$_{self}$, receipt ID qb4hg*). All the remaining parameters were identical to those employed within the *Autodock* protocol.

Note that we also submitted predictions using the same template-based protocol described in [16] (*HADDOCK$_{tb}$*; receipt ID nwm5a). Those led to the best performance from all our submissions but will not be discussed here.

### *Druggability calculations*

Following our previous study [13] we used the package f-pocket [40] to assess the druggability of the binding site within the ensembles of BACE-1 conformations generated by MD simulations. For each conformation, we evaluated the druggability score *D* [41] which ranges between 0 and 1 with higher values identifying more druggable geometries. It is customary to associate scores >0.5 to putative binding sites [41].

## Results and discussion

In the following we report the performance of our protocols in predicting near-native binding poses of BACE-1 ligands (stage 1a). Next we report the performance of Autodock in self-docking calculations of the same ligands onto the experimental structure of the receptor released at stage 1b. Evaluations were performed according to the data downloaded from the D3R website (https://drugdesigndata.org/about/grand-challenge-4-evaluation-results). The accuracy of the poses was evaluated by calculating the RMSD of each ligand with respect to its experimental reference structure (considering only heavy atoms), after superposition of the binding interface areas. First, we discuss the performance in terms of both global and per ligand descriptors. Next, to identify possible sources of errors, we analyze the accuracy of our EDES-like approach in sampling holo-like and druggable conformations of BACE-1, and the performance of the template-based algorithm to generate ligand conformers [16]. We conclude by summarizing the possible drawbacks of the methodology and its unique features, and by listing possible routes of future development.

### *Stage 1a*

Table 1 provides an overview of the performance of our methodologies in finding near-native poses of the 20 BACE-1 ligands. The best results were obtained with AutoDock [20] coupled to a multi-step structural optimization and rescoring of the top 10 poses (*Autodock$_{rr}$*). Using this approach, we found median and average values of the RMSD calculated on the top pose of each ligand (hereafter $\mathrm{RMSD}^{1}_{\mathrm{med}}$ and $\langle\mathrm{RMSD}^{1}\rangle$) lower than 2 Å and 3 Å respectively. Moreover, these values were lower than 1.5 Å and 2 Å when calculated on the nearest-native poses (hereafter $\mathrm{RMSD}^{\mathrm{min}}_{\mathrm{med}}$ and $\langle\mathrm{RMSD}^{\mathrm{min}}\rangle$, respectively). The AMBER-based refinement led to significant improvements with respect to the “standard” Autodock protocol, for which we obtained values of $\mathrm{RMSD}^{\mathrm{min}}_{\mathrm{med}}$ and $\langle\mathrm{RMSD}^{\mathrm{min}}\rangle$ lower than 2.5 Å. A very similar performance was achieved by HADDOCK when using standard settings, while the explicit consideration of non-polar hydrogen atoms of the ligand during docking led to an appreciable drop in the accuracy.

The $\langle\mathrm{RMSD}^{\mathrm{min}}\rangle$ and $\langle\mathrm{RMSD}^{1}\rangle$ metrics place our methods in the middle-left and middle-right regions of the histogram plot summarizing performances of all applicants (Fig. 2). An inspection of the protocols employed in this competition that perform better than any of ours (in terms of $\langle\mathrm{RMSD}^{1}\rangle$) revealed that our methodology is (among those for which these details were disclosed) the only one based on ensemble-docking from protein conformations generated *in silico* from an *apo* experimental structure of BACE-1. Thus, no information on the structures of its complexes with similar ligands was exploited to bias the conformation of the binding site towards holo-like geometries. In order to investigate more in detail the performance of our method, and possibly to correlate the accuracy of our results to one or more relevant parameters, we report the results for each of the 20 BACE ligands in Table 2, as well as in Figs. 3a, 3b, 3c, and 3d for *Autodock*, $Autodock_{rr}$, *HADDOCK*, and $HADDOCK_{all\text{-}Hs}$, respectively. The Table reveals that the aforementioned approaches gave at least one pose with $RMSD_{lig}$ < 2.5 Å respectively in 15, 16, 17, and 10 out of twenty cases, corresponding to success rates of 75%, 80%, 85%, and 50%.

In the following, we will discuss in more detail only the top three approaches: *Autodock*, $Autodock_{rr}$, and *HADDOCK*. Inspection of Table 2 reveals that the most challenging ligand was $BACE_{02}$, the only one for which we obtained poses featuring an RMSD > 3 Å from the native conformation with all approaches. Additional challenging ligands include $BACE_{10}$, for which the best RMSD value was 2.8 Å (obtained with $Autodock_{rr}$), and to a minor extent $BACE_{07}$, $BACE_{09}$, $BACE_{14}$, $BACE_{16}$, and $BACE_{18}$, for which 1 out of the three protocols was unable to find poses with RMSD values lower than 2.5 Å.

Inspection of Figs. 3a-b reveals the large variability in the orientation of docking poses for almost all of the ligands investigated. Such behavior, resulting in several poses displaying large RMSD values and in high standard deviations of the averages, was somewhat expected because of the (desired) tendency of our protocol to maximize the conformational diversity of the binding site among the protein structures used in docking calculations (*vide infra*). Clearly, considering the average RMSD over all poses would intrinsically penalize approaches like ours. Nonetheless, the best overall performing method ($Autodock_{rr}$) demonstrated its ability to reproduce at least one near-native pose among the top 5 for virtually all ligands but $BACE_{02}$. Allowing some degree of flexibility of the ligands (e.g. by activating torsional angles in AutoDock) could in principle improve results for this and the most challenging cases, although improvements were reported to be system-dependent [42, 43]. This is confirmed by the comparison with results obtained using HADDOCK, which includes by default flexibility of both docking partners by means of short MD runs in the space of the torsional angles.

Finally, we compared our predictions with docking calculations performed with Autodock on the experimental apo structure of BACE-1 (hereafter $Autodock_{apo}$). Table 1 confirms that, as expected, the lack of inclusion of protein flexibility has a major impact on the accuracy of near-native pose predictions [42–44]. In a further effort to identify most likely sources of errors we assessed the accuracies of our protocols in sampling near-native conformations of the protein and the ligands prior to docking calculations.

*Sampling of holo-like (and druggable) conformations of BACE-1*

The ability of our enhanced-sampling protocol to generate holo-like conformations of BACE-1 was evaluated in terms of the RMSD distributions calculated for the non-hydrogenous atoms of the binding site (hereafter $RMSD_{BS}$) with respect to each of the 20 BACE-1 experimental structures (provided at stage 1b) for $MD_{apo}$, $EDES_{3w}$ and the ensemble of 200 cluster structures used in docking calculations.

Fig. 4 shows that both $MD_{apo}$ and, to a larger extent, $EDES_{3w}$, were able to generate a significant fraction of receptor conformations displaying $RMSD_{BS}$ values lower than 2 Å with respect to every ligand/BACE-1 experimental structure. The good performance of $MD_{apo}$ is consistent with the relatively small conformational rearrangements undergone by BACE-1 upon binding of all ligands (Fig. 1c), corresponding to a decrease of $RoG_{BS}$ in the range 9.10-9.44 Å from the initial value of 9.79 Å found in 1SGZ. In agreement with previous results [13], $EDES_{3w}$ performs better than $MD_{apo}$ as testified by the sizeable shoulder seen in Fig. 4b, raising the percentage of structures with $RMSD_{BS}$ < 1.5 Å. Moreover, our multi-step cluster analysis confirmed its tendency to select a large (even larger than that sampled along the MD trajectories) fraction of low-$RMSD_{BS}$ geometries with respect to all the experimental reference structures (Fig. 4 and Table 3). It is worth pointing out that in all cases we obtained some conformations of the binding site that are virtually identical to the experimental structures, as testified by the lowest $RMSD_{BS}$ values, all around 1 Å.

We also evaluated the performance of our method in generating druggable conformations of BACE-1 [41]. Table 4 shows the results of this analysis on the 200 receptor conformations used for ensemble-docking calculations, compared with the values obtained for the targets investigated in [13]. It clearly appears that also the approach used in this work can generate a consistent fraction of structures associated with a large druggability score *D*.

*Generation of near-native ligand conformers*

The performance of our template-based similarity protocol in generating near-native conformations of the 20 BACE-1 ligands is summarized in Table 3. This Table reports the statistics of the RMSD calculated on heavy atoms of each ligand after structural alignment on the reference conformation extracted from the experimental structure of the corresponding complex (hereafter indicated as $RMSD_{lig\text{-}fit}$, to be distinguished from the same value calculated in the complex after alignment of the protein interface region). In all cases the minimum $RMSD_{lig\text{-}fit}$ values are lower than 2 Å, confirming the accuracy of the approach in reproducing near-native conformations [16]. Overall, these values are slightly larger than those obtained for the sampling of holo-like conformations of the receptor ($2_{nd}$ and $3_{rd}$ columns in Table 3). Moreover, in 4 (1) out of twenty cases we obtained an average $RMSD_{lig\text{-}fit}$ > 2 (2.5) Å, and in 6 out of twenty cases we obtained values of $RMSD^{min}_{lig\text{-}fit} > 1.5$ Å, and these ligands are (except for $BACE_{03}$ and $BACE_{15}$) exactly those for which we obtained the less accurate docking results. This further confirms previous findings on the importance of sampling not only the correct conformations of the protein but also that of the ligands, as well as the exponential impact on accuracy arising from the combination of even minimal displacements from the correct structures in both partners [16].

*Stage1b*

This is a self-docking stage for which the bound protein structures – but not those of the compounds – are known. The comparison between the data obtained at stages 1a and 1b is instructive to evaluate the performance of the docking protocol in the presence of the correct structures of the receptor. This enables to highlight drawbacks likely unrelated to the protein flexibility problem, as well as to confirm the importance of using a relevant fraction of correct conformers of the ligand in ensemble-docking calculations. We performed this exercise using AutoDock; namely, we docked each of the 20 ligands on the corresponding experimental structure of the receptor.

Table 5 reports the overall performance of $Autodock_{self}$, while Fig. 3e shows a more detailed analysis of the top 5 poses for each BACE-1 ligand. Interestingly, a marginal improvement was seen with respect to the results obtained with the *Autodock* protocol at stage 1a, while the success rate evaluated as the number of ligands for which at least one pose featured a value of the RMSD ≤ 2.5 Å remained 75%. This result was in part expected, in view of the good sampling of holo-like protein conformations obtained with our modified EDES protocol (Table 3). Moreover, also in this case the most challenging ligands include those featuring the largest values of $\langle RMSD_{lig\text{-}ref}\rangle$ and/or $RMSD^{min}_{lig\text{-}ref}$ in Table 3. In particular, wrong poses were found for $BACE_{09}$, $BACE_{10}$, $BACE_{15}$, $BACE_{16}$, and $BACE_{18}$, while for $BACE_{07}$ all poses have RMSD values close to 2.5 Å. While further relaxation and rescoring of these poses is expected to increase the success rate, we note that a very minor conformational change towards the correct geometry of the binding site was sufficient to find at least one pose with RMSD lower than 2 Å for $BACE_{02}$.

# Conclusions and Perspectives

We report the performance of our hybrid ensemble-docking approach in its first participation to a D3R Grand Challenge competition. The approach is founded on a template-based algorithm to select proper ligand conformers, and on our recently published EDES protocol (implemented here with small modifications with respect to the original version) to sample holo-like protein conformations starting from the apo one. EDES confirmed its excellent performance in sampling holo-like conformations of the protein (particularly at the binding site) for all of the twenty complexes formed between BACE-1 and the congeneric ligands subject of this study. A very good accuracy in reproducing near-native ligand conformers was achieved also by our template-based approach. These performances reflected in the relatively high accuracy in the prediction of near-native binding poses. Independently of the docking program used, our method was able to find near-native poses among the top 5 ones for at least 75% of the twenty complexes subject of the pose prediction sub-challenge. While HADDOCK found near-native poses for more targets than AutoDock, the latter featured the best overall performance when coupled to a computationally cheap post-docking relaxation of the poses. Performing docking calculations on the apo experimental structure of BACE-1 resulted in significantly less accurate predictions, due to the unaccounted rearrangements of the protein flap occurring upon ligand binding. Finally, the good performance of our approach was testified by the only slight overall improvement obtained when performing self-docking calculations on the twenty experimental holo receptor structures.

Note that in this Grand Challenge we also submitted pose predictions following the template-based protocol described in [16] ($HADDOCK_{tb}$) in which an ideal choice of both ligand and receptor conformations is made based on ligand similarity to known PDB entries. As in GC3, this approach led to an excellent performance with $\langle RMSD^{min}\rangle$, $\langle RMSD^{1}\rangle$ and $\langle RMSD\rangle$ respectively of 1.4 Å, 1.66 Å and 1.95 Å, which demonstrates that a template-based approach remains the best strategy when 3D structures of related complexes are available in the PDB. However, when this is not the case, our EDES approach appears to be an attractive alternative. Moreover, it has been proposed that including MD-generated receptor conformations (such as those obtained from EDES) in the virtual screening protocol could promote the discover of new active chemotypes, especially for flexible receptors, in which entirely new pocket conformations may be revealed for potential ligand binding [45–47].

Further developments of the method will include improved identification of putative binding sites and of their key residues, coupling of EDES with the use of co-solvents, and exploitation of different cluster methodologies.

# Acknowledgments

A.B. gratefully acknowledges the Sardinia Regional Government for the financial support of his Ph.D. scholarship (P.O.R. Sardegna F.SE., Operational Programme of the Autonomous Region of Sardinia, European Social Fund 2014–2020—Axis III Education and Training, Thematic Goal 10, Priority of Investment 10ii, Specific Goal 10.5., Action Partnership Agreement 10.5.12). This work was done as part of the BioExcel CoE (www.bioexcel.eu), a project funded by the European Union Horizon 2020 Program under Grant Agreements 675728 and 823830 (to A. M. J. J. B.) with financial support from the Dutch Foundation for Scientific Research (NWO) (TOP-PUNT grant 718.015.001.

# Tables

**Table 1.** Overall performance of our protocols in retrieving near-native ligands conformations of BACE-1 ligands (rows 4 to 8) during stage 1a, and performance of the *Autodock$_{self}$* protocol in stage 1b (last row; data from https://drugdesigndata.org). The values in parentheses are standard deviations of the averages.

| | **Averages** | | | **Median** | | |
|---|---|---|---|---|---|---|
| **Protocol** | **$\langle RMSD^{min} \rangle$** | **$\langle RMSD^{1} \rangle$** | **$\langle RMSD \rangle$** | **$RMSD^{min}_{med}$** | **$RMSD^{1}_{med}$** | **$RMSD_{med}$** |
| **Stage 1a** | | | | | | |
| ***Autodock$_{rr}$*** | 1.73 (0.88) | 2.86 (2.71) | 4.24 (1.77) | 1.38 | 1.78 | 3.89 |
| ***Autodock*** | 2.48 (1.82) | 3.10 (2.57) | 4.41 (2.10) | 2.07 | 2.25 | 4.28 |
| ***HADDOCK*** | 2.28 (0.99) | 4.12 (2.73) | 4.64 (1.53) | 2.06 | 3.12 | 4.23 |
| ***HADDOCK$_{all\text{-}Hs}$*** | 3.19 (2.26) | 4.83 (3.50) | 5.96 (2.18) | 2.66 | 3.10 | 5.76 |
| ***Autodock$_{apo}$*** | 3.78 (2.94) | 5.67 (3.72) | 5.17 (3.34) | 2.47 | 3.51 | 3.49 |
| **Stage 1b** | | | | | | |
| ***Autodock$_{self}$*** | 2.24 (2.13) | 2.93 (2.78) | 3.59 (2.73) | 1.60 | 2.03 | 2.30 |

**Table 2.** Summary the docking results obtained with the four methods described in this work for each of the 20 BACE-1 ligands (data from https://drugdesigndata.org). $RMSD^{min}$ values larger than 2.5 Å are bolded.

| | ***Autodock*** | | | ***Autodock$_{rr}$*** | | | ***HADDOCK*** | | | ***HADDOCK$_{all\text{-}Hs}$*** | | |
|---|---|---|---|---|---|---|---|---|---|---|---|---|
| **Target Ligand** | **$RMSD^{min}$** | **$RMSD^{1}$** | **$\langle RMSD \rangle$** | **$RMSD^{min}$** | **$RMSD^{1}$** | **$\langle RMSD \rangle$** | **$RMSD^{min}$** | **$RMSD^{1}$** | **$\langle RMSD \rangle$** | **$RMSD^{min}$** | **$RMSD^{1}$** | **$\langle RMSD \rangle$** |
| $BACE_{01}$ | 1.7 | 2.2 | 3.7 (2.9) | 1.2 | 1.2 | 1.8 (0.5) | 1.5 | 2.0 | 2.8 (1.7) | **3.1** | 3.1 | 4.7 (2.7) |
| $BACE_{02}$ | **4.2** | 4.2 | 6.5 (2.7) | **4.5** | 4.5 | 7.9 (2.9) | **3.9** | 4.4 | 4.3 (0.3) | **10.2** | 10.5 | 11.0 (0.7) |
| $BACE_{03}$ | **2.8** | 2.8 | 5.3 (3.4) | 1.8 | 2.6 | 3.8 (3.4) | 2.5 | 3.5 | 4.2 (2.7) | **3.2** | 3.5 | 5.7 (3.2) |
| $BACE_{04}$ | 1.5 | 1.5 | 2.9 (1.3) | 1.1 | 1.1 | 2.2 (1.0) | 2.2 | 9.6 | 6.6 (4.0) | 2.3 | 9.7 | 7.1 (3.6) |
| $BACE_{05}$ | 2.1 | 2.1 | 3.8 (3.3) | 1.4 | 1.6 | 5.3 (4.9) | 1.4 | 3.1 | 3.8 (3.1) | **8.7** | 9.8 | 10.0 (0.9) |
| $BACE_{06}$ | 1.1 | 1.1 | 1.7 (0.7) | 1.5 | 1.6 | 2.1 (0.8) | 1.8 | 2.3 | 3.6 (3.3) | 2.3 | 2.3 | 5.8 (4.6) |
| $BACE_{07}$ | 2.1 | 2.5 | 2.5 (0.3) | **2.6** | 2.6 | 3.4 (0.9) | 2.4 | 2.6 | 4.5 (3.2) | **3.5** | 3.6 | 6.2 (3.5) |
| $BACE_{08}$ | 1.2 | 1.7 | 3.3 (4.0) | 1.0 | 1.0 | 4.9 (5.0) | 1.3 | 1.6 | 2.0 (1.1) | 1.4 | 1.5 | 5.0 (4.7) |
| $BACE_{09}$ | 2.2 | 2.2 | 4.5 (2.9) | 2.4 | 3.1 | 4.1 (3.1) | **3.3** | 10.0 | 8.4 (3.0) | 2.2 | 2.9 | 5.3 (3.8) |
| $BACE_{10}$ | **9.3** | 9.4 | 9.4 (0.1) | **2.8** | 10.0 | 8.2 (3.0) | **5.5** | 5.8 | 6.6 (2.0) | **3.5** | 9.5 | 6.5 (2.9) |
| $BACE_{11}$ | 1.5 | 2.8 | 2.3 (0.8) | 1.0 | 1.2 | 3.6 (4.3) | 2.0 | 4.5 | 3.7 (1.0) | 1.7 | 1.7 | 5.4 (3.9) |
| $BACE_{12}$ | 1.4 | 1.5 | 1.8 (0.5) | 1.2 | 10.3 | 6.8 (4.7) | 1.2 | 1.9 | 3.5 (3.4) | 1.4 | 1.9 | 4.8 (4.3) |
| $BACE_{13}$ | 1.5 | 2.2 | 2.1 (0.5) | 0.9 | 1.4 | 3.6 (4.2) | 1.5 | 1.5 | 3.9 (3.2) | 1.8 | 1.8 | 2.2 (0.4) |
| $BACE_{14}$ | **4.2** | 10.8 | 8.2 (3.6) | 1.3 | 2.5 | 4.9 (3.3) | 2.1 | 9.4 | 6.9 (3.8) | **2.9** | 3.1 | 4.7 (3.5) |
| $BACE_{15}$ | 2.2 | 2.2 | 4.4 (3.8) | 1.1 | 1.8 | 2.4 (1.0) | 2.2 | 3.0 | 3.7 (1.6) | **3.4** | 9.4 | 8.5 (2.9) |
| $BACE_{16}$ | 2.5 | 2.5 | 4.2 (3.6) | **2.6** | 2.6 | 4.4 (2.9) | 2.5 | 3.1 | 4.5 (3.5) | **3.0** | 3.3 | 6.0 (4.0) |
| $BACE_{17}$ | 1.8 | 1.9 | 4.8 (2.7) | 1.4 | 1.7 | 3.1 (2.4) | 1.9 | 5.7 | 5.2 (1.9) | 1.7 | 2.3 | 3.0 (1.9) |
| $BACE_{18}$ | **2.6** | 2.6 | 5.3 (2.5) | 1.8 | 1.8 | 4.7 (2.5) | 2.0 | 3.9 | 4.0 (1.4) | 1.9 | 2.4 | 3.5 (2.0) |
| $BACE_{19}$ | 1.8 | 2.3 | 6.8 (4.4) | 1.3 | 1.5 | 3.4 (3.0) | 2.3 | 2.3 | 5.1 (3.3) | 2.4 | 9.6 | 8.1 (3.2) |
| $BACE_{20}$ | 2.1 | 2.1 | 4.9 (3.4) | 1.6 | 8.5 | 4.0 (3.0) | 2.0 | 2.0 | 5.3 (4.0) | **3.2** | 3.2 | 5.9 (3.6) |

**Table 3.** Performances of our methodology evaluated separately for the generation of protein and ligand conformations similar to those found in the ligand/BACE-1 experimental structures. The 2nd column reports the lowest $RMSD_{BS}$ calculated across the 200 receptor conformations with respect to each experimental structure. The 3rd column reports the percentage of conformations displaying an $RMSD_{BS}$ lower than 1.5 Å. The last column reports the minimum and maximum RMSD values (calculated on the non-hydrogenous atoms with respect to the structure of each ligand in the experimental structure), as well as the average and standard deviation within parentheses. Values of $RMSD_{lig\text{-}ref}^{min}$ larger than 1.5 Å and average values of $RMSD_{lig\text{-}ref}$ larger than 2 Å are underlined and bolded, respectively.

| | **Protein** | | **Ligands** |
|---|---|---|---|
| **System** | **$RMSD_{BS}^{min}$ [Å]** | **% $RMSD_{BS}$ < 1.5 Å** | **$RMSD_{lig\text{-}ref}$ [Å]** |
| BACE$_{01}$ | 1.13 | 16 | 1.11-1.71 (1.44±0.24) |
| BACE$_{02}$ | 1.08 | 15 | **1.16-3.08 (2.70±0.63)** |
| BACE$_{03}$ | 1.07 | 15 | 0.95-1.44 (1.25±0.15) |
| BACE$_{04}$ | 1.08 | 19 | 1.03-2.05 (1.55±0.44) |
| BACE$_{05}$ | 1.07 | 15 | 0.82-3.05 (1.50±0.73) |
| BACE$_{06}$ | 1.19 | 15 | 0.58-1.38 (0.98±0.23) |
| BACE$_{07}$ | 1.10 | 17 | **1.62-2.77 (2.08±0.45)** |
| BACE$_{08}$ | 1.07 | 17 | 0.57-1.51 (0.92±0.29) |
| BACE$_{09}$ | 1.17 | 17 | 1.52-2.38 (1.96±0.38) |
| BACE$_{10}$ | 1.13 | 17 | 1.54-2.33 (1.98±0.33) |
| BACE$_{11}$ | 1.06 | 17 | 1.03-2.24 (1.64±0.48) |
| BACE$_{12}$ | 1.07 | 17 | 0.68-1.79 (1.13±0.44) |
| BACE$_{13}$ | 1.06 | 16 | 0.52-1.00 (0.75±0.16) |
| BACE$_{14}$ | 0.98 | 16 | **1.37-3.34 (2.48±0.81)** |
| BACE$_{15}$ | 1.16 | 14 | 1.53-2.23 (1.95±0.33) |
| BACE$_{16}$ | 1.17 | 14 | **1.83-3.53 (2.39±0.59)** |
| BACE$_{17}$ | 1.13 | 14 | 1.08-1.58 (1.34±0.18) |
| BACE$_{18}$ | 1.14 | 14 | 1.53-2.01 (1.81±0.11) |
| BACE$_{19}$ | 1.07 | 16 | 1.17-1.51 (1.33±0.10) |
| BACE$_{20}$ | 1.20 | 15 | 1.49-3.26 (1.83±0.52) |

**Table 4.** Performance of our approach in generating druggable conformations of the binding site. The percentages of structures featuring druggability scores *D* larger than 0.5 to 0.9 are reported in columns 2 to 6, respectively.

| **% structures with *D* greater than** | **0.5** | **0.6** | **0.7** | **0.8** | **0.9** |
|---|---|---|---|---|---|
| **BACE** | 12.5 | 8.0 | 6.5 | 2.5 | 1.0 |
| **BGT*** | 15.8 | 10.2 | 6.8 | 4.6 | 1.4 |
| **RIC*** | 3.2 | 2.4 | 1.6 | 0.4 | 0.2 |
| **ABP*** | 7.6 | 5.2 | 3.0 | 1.8 | 0.4 |

*From [13]. BGT: T4 phage β-glucosyltransferase; RIC: recombinant ricin; ABP: allose binding protein.

**Table 5.** Summary of the self-docking results obtained with the *Autodock*$_{self}$ protocol for each of the 20 BACE-1 ligands. $RMSD^{min}$ values larger than 2.5 Å are bolded.

| **Target Ligand** | **$RMSD^{min}$** | **$RMSD^{1}$** | **⟨RMSD⟩** |
|---|---|---|---|
| $BACE_{01}$ | 1.5 | 1.7 | 4.8 (4.5) |
| $BACE_{02}$ | 1.7 | 3.5 | 5.1 (3.1) |
| $BACE_{03}$ | 1.8 | 2.4 | 2.0 (0.3) |
| $BACE_{04}$ | 1.0 | 1.1 | 1.1 (0.1) |
| $BACE_{05}$ | 0.9 | 0.9 | 1.2 (0.5) |
| $BACE_{06}$ | 1.1 | 1.1 | 1.4 (0.4) |
| $BACE_{07}$ | 2.3 | 2.3 | 2.4 (0.1) |
| $BACE_{08}$ | 0.8 | 1.1 | 0.9 (0.2) |
| $BACE_{09}$ | **9.7** | 10.1 | 10.2 (0.5) |
| $BACE_{10}$ | **3.2** | 9.9 | 8.6 (3.0) |
| $BACE_{11}$ | 1.0 | 1.0 | 2.4 (1.3) |
| $BACE_{12}$ | 0.8 | 1.1 | 1.2 (0.3) |
| $BACE_{13}$ | 0.7 | 0.9 | 0.9 (0.1) |
| $BACE_{14}$ | 1.9 | 1.9 | 5.7 (3.4) |
| $BACE_{15}$ | **3.3** | 3.3 | 5.4 (1.8) |
| $BACE_{16}$ | **3.3** | 3.8 | 6.6 (4.2) |
| $BACE_{17}$ | 1.7 | 2.0 | 2.1 (0.3) |
| $BACE_{18}$ | **5.6** | 5.6 | 5.7 (0.1) |
| $BACE_{19}$ | 1.2 | 2.0 | 2.0 (0.4) |
| $BACE_{20}$ | 1.5 | 2.3 | 2.2 (0.5) |

# Figures

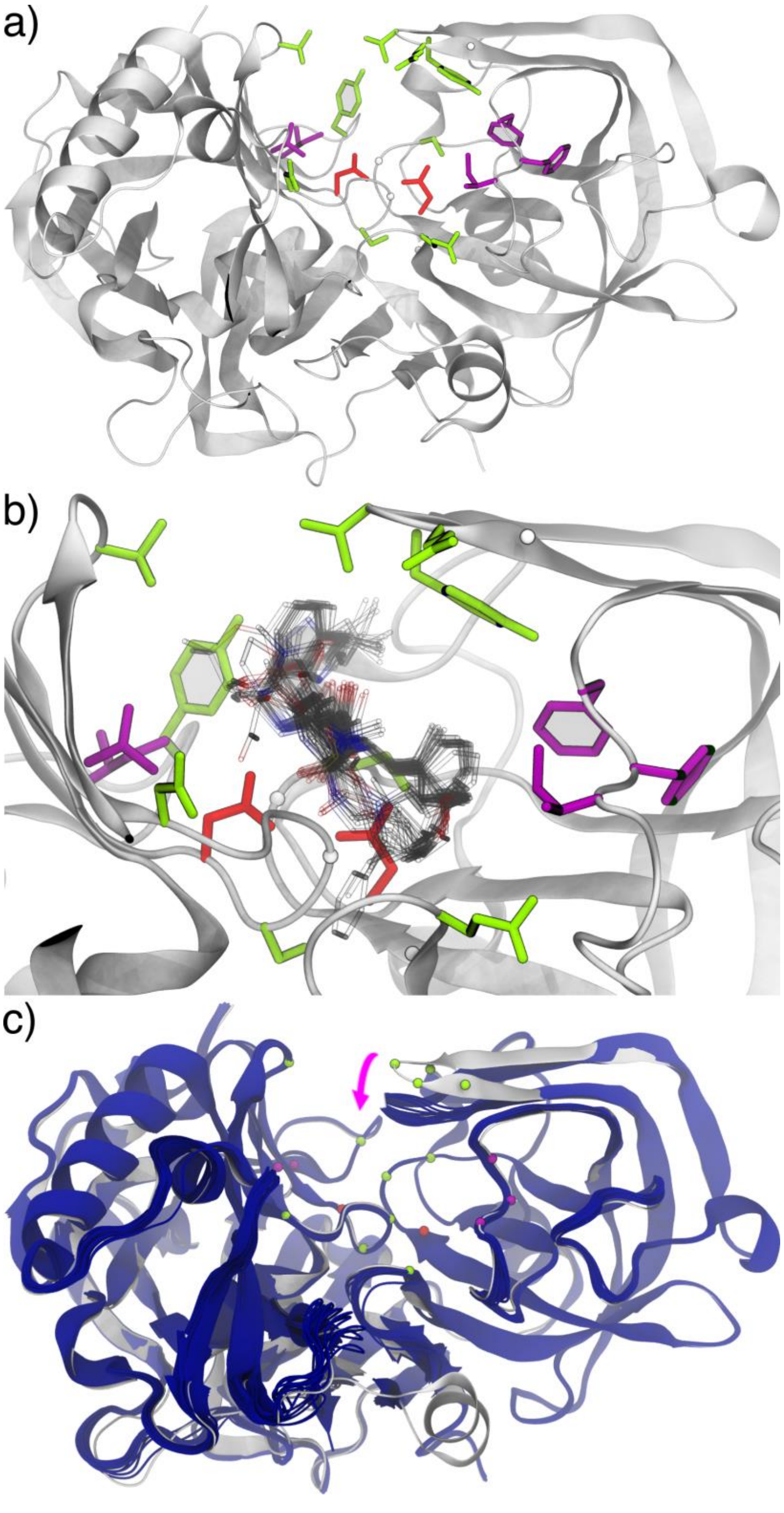

**Fig. 1** Putative binding site identified on the BACE-1 apo protein (PDB ID 1SGZ [17]) for implementation of the EDES approach. a) Structure of the protein (grey ribbons) showing the sidechains of the 20 residues in Table 2 as sticks colored by type (polar, apolar, acidic and glycines in light green, magenta, red and white respectively); b) zoom on the putative binding site in a), showing in transparent sticks the experimental poses of the 20 ligands provided by the organizers (after superposition of common $C_\alpha$ atoms on all proteins to 1SGZ); c) comparison between the apo structure of BACE-1 (grey ribbons) and the 20 ligand/BACE-1 complex structures (blue ribbons) released at stage 1b of the challenge. The magenta arrow highlights the major displacement undergone by the protein flap upon ligand binding.

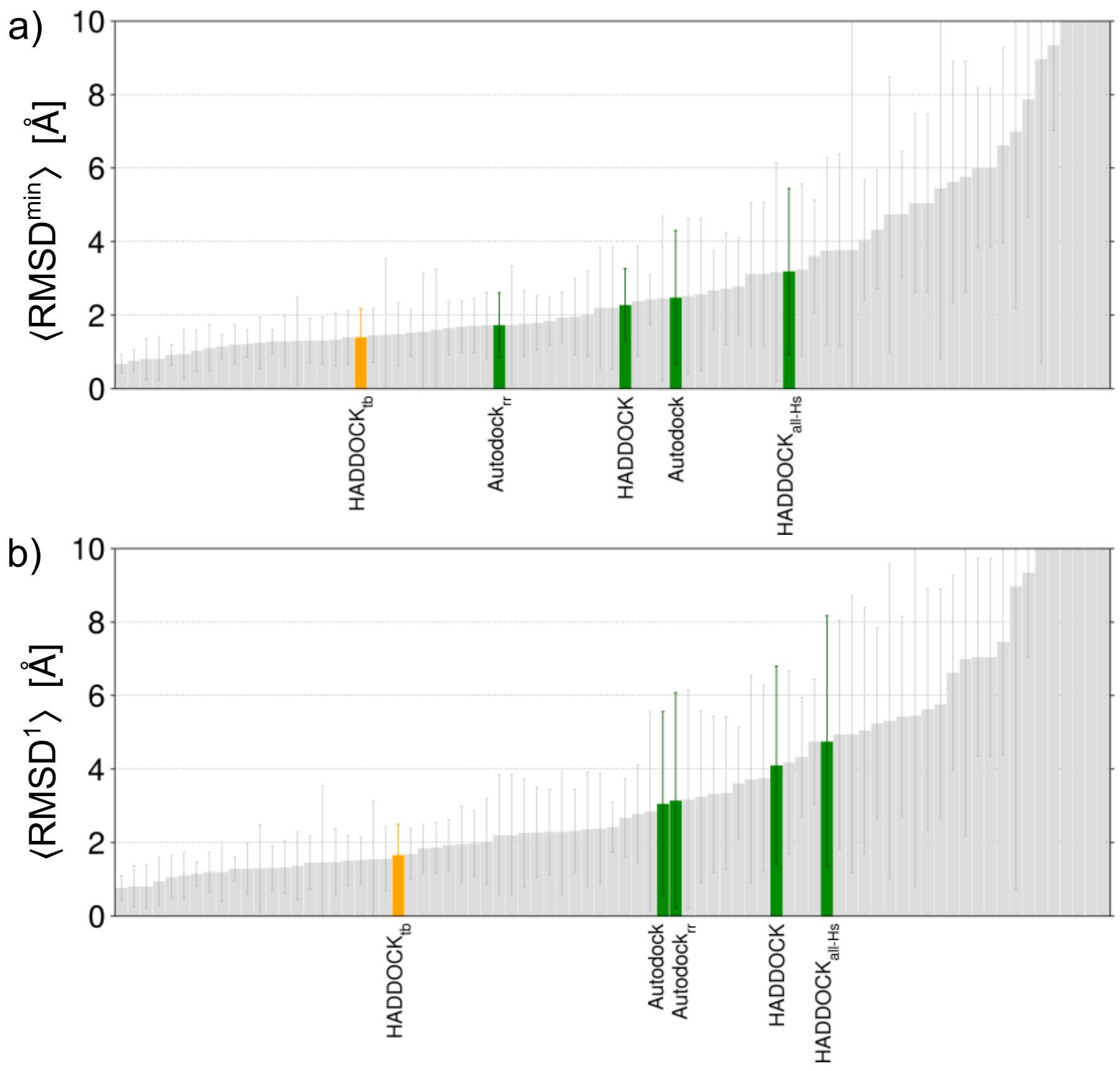

**Fig. 2** Overall performance of the protocols employed in this study, as measured by the values of $\langle \mathrm{RMSD}^{\mathrm{min}} \rangle$ and $\langle \mathrm{RMSD}^{1} \rangle$

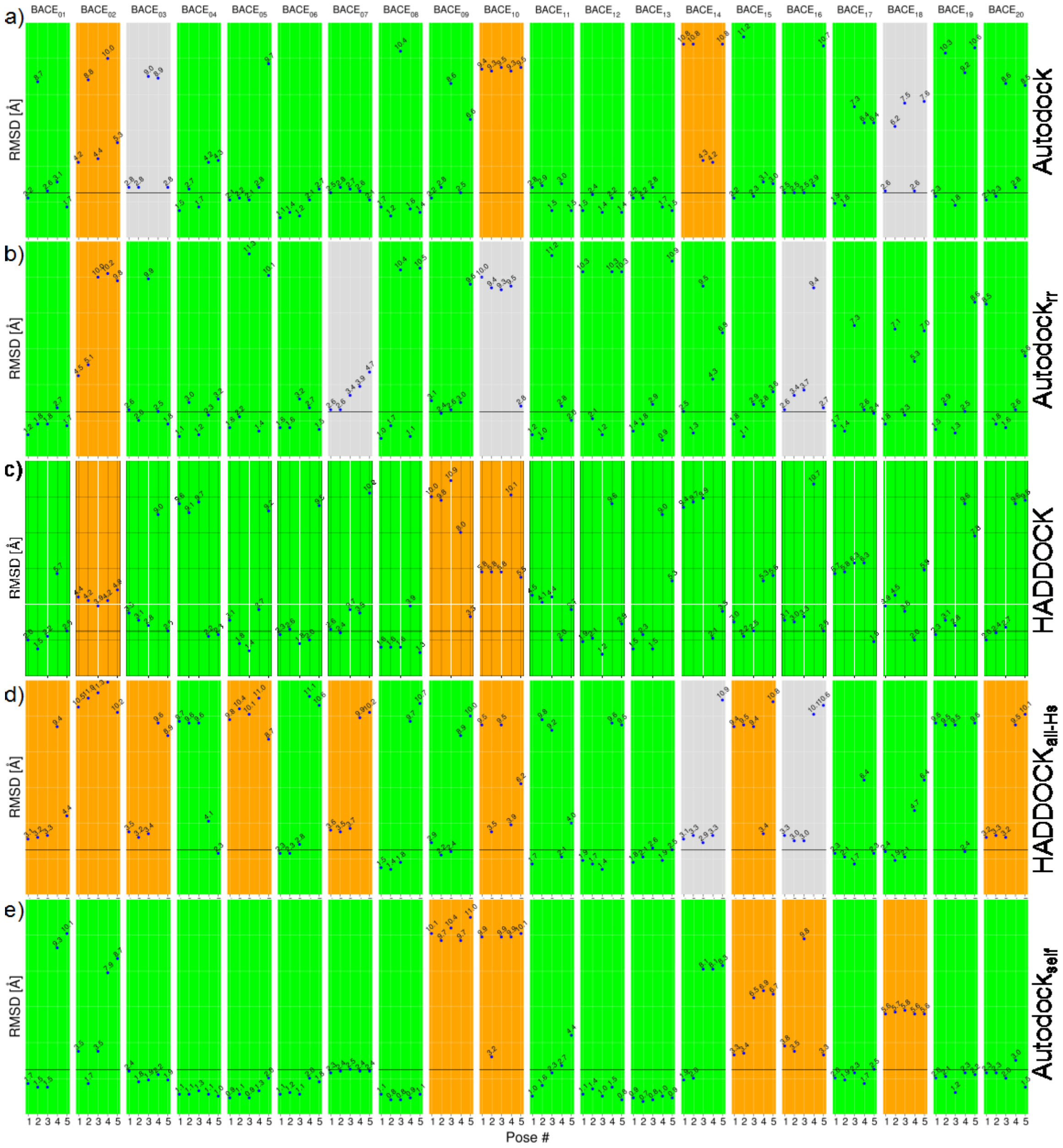

**Fig. 3** Performance of the Autodock (a), Autodock$_{rr}$ (b), HADDOCK (c) and HADDOCK$_{all\text{-}Hs}$ (d) protocols in reproducing the near-native conformations of the 20 BACE-1 ligands. Green and grey panels refer to targets for which we obtained at least one pose within the top 5 featuring a value of the ligand RMSD ≤ 2.5 Å and ≤ 3 Å respectively, while orange boxes indicate cases for which no such poses were found among the top 5 ones

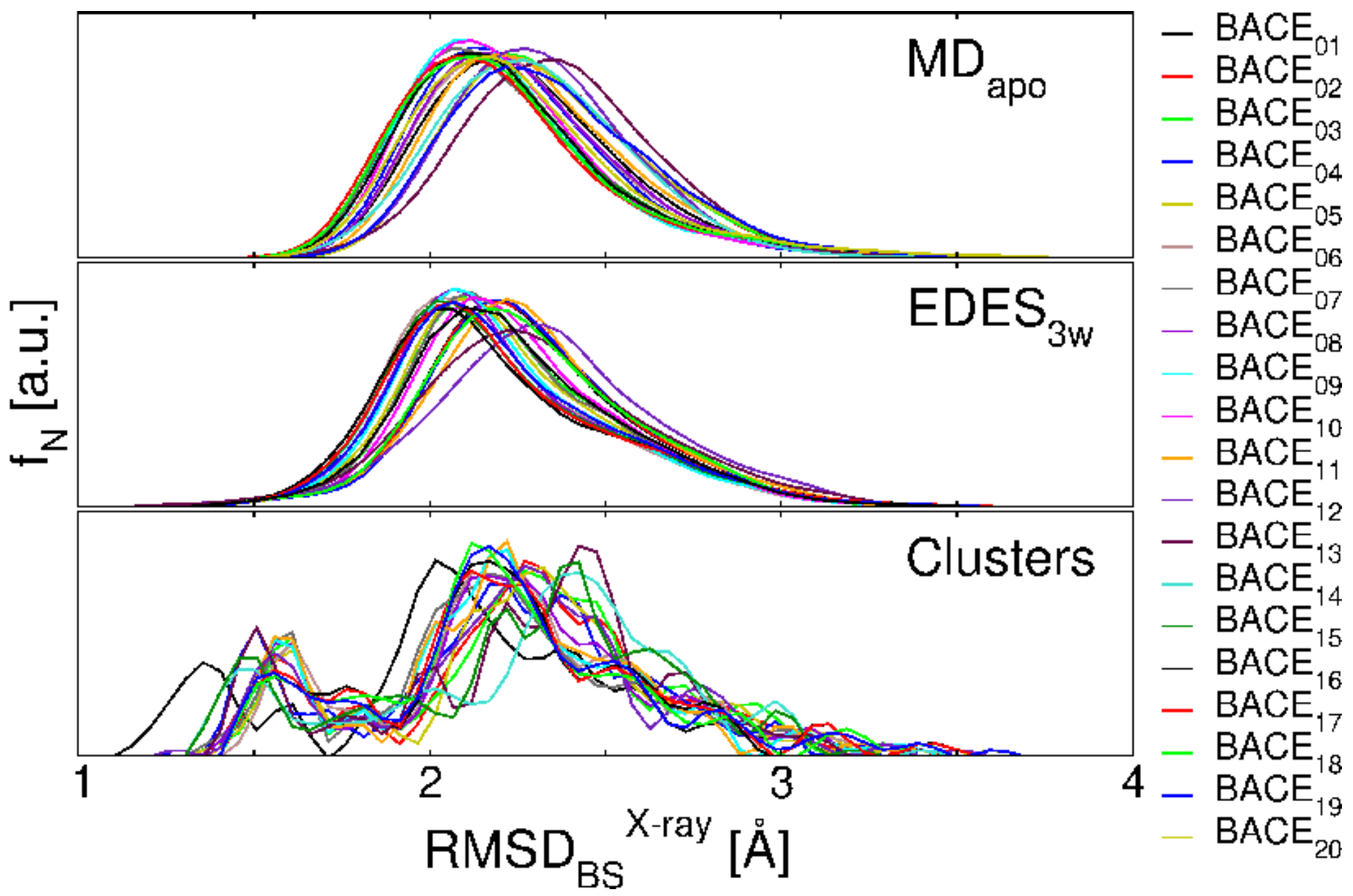

**Fig. 4** Normalized distributions (bin size = 0.1 Å) of $RMSD_{BS}$ calculated with respect to the 20 experimental structures of ligands in complex with BACE-1 for $MD_{apo}$ (upper graph) and $EDES_{3w}$ (middle panel) trajectories, as well as for the ensemble of 200 BACE-1 structures used in ensemble docking calculations (lower panel)

# Supplementary Tables

**Table S1** Ligand templates structures. For each target compound (first column), we report the template PDB ID (second column) and the Tanimoto similarity coefficient between the two ligands (third column). The Tanimoto coefficient ranges from 0 (lowest similarity) to 1 (highest similarity).

| Ligand target | Template PDB ID | Tanimoto similarity |
|---|---|---|
| $BACE_{01}$ | 3DV1 | 0.605 |
| $BACE_{02}$ | 3DV1 | 0.875 |
| $BACE_{03}$ | 3DV1 | 0.821 |
| $BACE_{04}$ | 3DV1 | 0.872 |
| $BACE_{05}$ | 3DV1 | 0.725 |
| $BACE_{06}$ | 4DPI | 0.660 |
| $BACE_{07}$ | 2IQG | 0.618 |
| $BACE_{08}$ | 3DV5 | 0.543 |
| $BACE_{09}$ | 3DV5 | 0.698 |
| $BACE_{10}$ | 3K5C | 0.739 |
| $BACE_{11}$ | 3VEU | 0.833 |
| $BACE_{12}$ | 4KE1 | 0.681 |
| $BACE_{13}$ | 3K5C | 0.681 |
| $BACE_{14}$ | 3K5C | 0.861 |
| $BACE_{15}$ | 3K5C | 0.891 |
| $BACE_{16}$ | 3K5C | 0.750 |
| $BACE_{17}$ | 2B8L | 0.490 |
| $BACE_{18}$ | 4R92 | 0.476 |
| $BACE_{19}$ | 3DV1 | 0.625 |
| $BACE_{20}$ | 6BFD | 0.604 |

**Table S2** List of residues defining the putative binding site of BACE-1 ligands investigated in this work, along with their occurrence frequencies in the list of residues within 3.5 Å of the ligands in the experimental structures with PDB IDs 3DV1, 4DPI, 2IQG, 3DV5, 3K5C, 3VEU, 4KE1, 2B8L, 4R92, 6BFD.

| Residue | Occurrence |
|---|---|
| Q12 | 3 |
| G13 | 1 |
| D32 | 9 |
| G34 | 9 |
| S35 | 2 |
| Y71 | 5 |
| T72 | 9 |
| Q73 | 7 |
| G74 | 2 |
| F108 | 4 |
| F109 | 1 |
| I110 | 1 |
| Y198 | 1 |
| I226 | 2 |
| D228 | 9 |
| S229 | 2 |
| G230 | 9 |
| V232 | 8 |
| N233 | 1 |
| T329 | 1 |